\documentclass{aa}

\usepackage{graphicx}
\usepackage{txfonts}
\usepackage{comment} 
\usepackage[colorlinks=true, citecolor=blue, linkcolor=blue]{hyperref}
\usepackage{natbib}
\begin{document} 

\bibpunct{(}{)}{;}{a}{}{,} 

   \title{AMICO galaxy clusters in KiDS-1000: Cosmological constraints from the two-point correlation function}
    
   \author{M. Romanello
          \inst{1,2}
          \and
          G. F. Lesci\inst{1,3}
          \and
          F. Marulli\inst{1,2,4}
          \and
          L. Moscardini\inst{1,2,4}
          \and
          C. Giocoli\inst{1,2,4}
          \and
          M. Maturi\inst{5,6}
          \and
          M. Radovich\inst{7}
          \and
          S. Bardelli\inst{2}
          \and
          G. Castignani\inst{2} 
          \and
          H. Hildebrandt\inst{8}
          \and 
          L. Ingoglia\inst{3}
          \and
          S. Joudaki\inst{9}
          \and
          E. Puddu\inst{10}
          \and
          M. Sereno\inst{2,4}
          }

   \institute{Dipartimento di Fisica e Astronomia “A. Righi” - Alma Mater Studiorum Università di Bologna, via Piero Gobetti 93/2, 40129 Bologna, Italy
   \and
   INAF - Osservatorio di Astrofisica e Scienza dello Spazio di Bologna, via Piero Gobetti 93/3, 40129 Bologna, Italy 
   \and 
   INAF - Istituto di Radioastronomia, Via Piero Gobetti 101, 40129, Bologna, Italy
   \and
   INFN - Sezione di Bologna, Viale Berti Pichat 6/2, 40127 Bologna, Italy
   \and 
   Zentrum für Astronomie, Universität Heidelberg, Philosophenweg 12, 69120 Heidelberg, Germany
   \and 
   ITP, Universität Heidelberg, Philosophenweg 16, 69120 Heidelberg, Germany
   \and 
   INAF - Osservatorio Astronomico di Padova, vicolo dell’Osservatorio 5, 35122 Padova, Italy
   \and
   Ruhr University Bochum, Faculty of Physics and Astronomy, Astronomical Institute (AIRUB), German Centre for Cosmological Lensing, 44780 Bochum, Germany
   \and
   Centro de Investigaciones Energéticas, Medioambientales y Tecnológicas (CIEMAT), Av. Complutense 40, E-28040 Madrid, Spain
   \and
   INAF - Osservatorio Astronomico di Capodimonte, Salita Moiariello 16, 80131 Napoli, Italy
\\\\
        \email{massimilia.romanell2@unibo.it}
             }


 
  \abstract
  {In this paper we study the clustering properties of an optically identified photometric galaxy cluster catalogue from the fourth data release of the Kilo Degree Survey (KiDS-1000). The sample was identified by the Adaptive Matched Identifier of Clustered Objects (AMICO) algorithm and consists of 9049 clusters with $S/N\geq3.5$, in the redshift range $0.1\leq z<0.8$, spread over an effective area of 838.78 deg$^2$. Clusters are selected according to their observed intrinsic richness, with thresholds of $\lambda^*_\mathrm{obs}\geq 20,25,30$ within $0.10\leq z<0.30$, $0.30\leq z<0.45$ and $0.45\leq z<0.80$, respectively. We measure the multipoles of the two-point correlation function over different ranges of scales from 5 $h^{-1}$ Mpc to 150 $h^{-1}$ Mpc. We model the clustering signal in the context of the $\Lambda$-cold dark matter ($\Lambda$CDM) cosmology, accounting for the main observational effects, such as geometric- and redshift-space distortions (RSDs), and photo-$z$ uncertainties. The effective bias of the cluster sample is estimated via a mass-richness scaling relation. For the Markov Chain Monte Carlo (MCMC) analysis, we adopt Gaussian priors on the parameters of the mass-richness scaling relation, and broad, uniform priors to derive cosmological constraints on the cold matter density parameter, the power spectrum amplitude and the structure growth parameter, $S_8\equiv\sigma_8 \sqrt{\Omega_\mathrm{m}/0.3}$. Specifically, we find $\Omega_\mathrm{m}=\Omega_\mathrm{CDM}+\Omega_\mathrm{b}=0.24^{+0.02}_{-0.02}$ and $\sigma_8=0.86^{+0.13}_{-0.10}$ representing an improvement in precision over both the 2D and 3D AMICO clustering analyses performed on the previous KiDS data release. Moreover, we obtain $S_8=0.77^{+0.10}_{-0.08}$ in $1\sigma$ agreement with the AMICO KiDS-1000 number counts constraints estimated from the same catalogue, as well as with some results from external early- and late-Universe data sets. Our work strengthens the potential of cluster clustering as a complementary and competitive probe, to be included and further developed with the final KiDS-Legacy data release, and in other ongoing and future cluster cosmological analyses based on Stage-IV photometric redshift surveys.}


   \maketitle
%

\section{Introduction} 
In the hierarchical cosmological framework, galaxy clusters are the latest gravitationally bound systems ever formed \citep[][]{Kaiser84}, emerging from the nodes of the filamentary cosmic web and tracing the peaks of the large-scale matter density field  \citep{Bardeen86, Tormen98}. As multi-component systems, made up of dark and baryonic matter in different phases, they provide valuable insights about the structure formation process along cosmic time. Their assembly history reveals information about the dynamical environment in which they grow, contributing to shed light on some of the major open questions in cosmology. These include the nature of dark matter \citep{Meneghetti20, Eckert22, Natarajan26} and dark energy \citep{Allen11, Ghirardini24}. \\
\indent Recently, numerous theoretical and observational studies have shown that the abundance and the spatial distribution of galaxy clusters are powerful tools for constraining fundamental parameters of the $\Lambda$-cold dark matter ($\Lambda$CDM) model, such as $\Omega_m$, the total matter density parameter, and $\sigma_8$, the square root of the mass variance on scales of 8 $h^{-1}$ Mpc \citep[see, for instance,][]{Costanzi19, Lesci_counts, Lesci22, Chiu23, Bocquet24, Fumagalli24, Ghirardini24, Romanello24, Sunayama24, Abbott25, Lesci25}. \\
\indent In particular, the two-point correlation function of galaxy clusters offers several significant advantages in cosmological inference. Firstly, compared to galaxies, they are highly biased tracers, so their clustering signal is higher. They are also less affected by nonlinearities, enabling us to simplify theoretical modelling within the same range of scales under investigation. In addition, the `fingers-of-God' (FoG) effect — the small-scale redshift-space distortions (RSDs) caused by the motions of tracers within virialised structures — is reduced \citep[see, for example,][]{Veropalumbo14, Sereno15, Marulli17}. Finally, galaxy clusters are identifiable in optical and near-infrared wavebands due to their luminous starlight \citep{Rykoff14, Rykoff16, Maturi19, Maturi25}; in X-rays by exploiting the emission from their diffuse intracluster medium \citep[ICM,][]{Pacaud16, Bulbul24}; and in the millimetric band through the Sunyaev-Zel’dovich effect \citep{Sunyaev70, Sunyaev72}, which is produced by the inverse Compton scattering between the hot ICM electrons and the cosmic microwave background (CMB) photons \citep{Planck11, Hilton21}. This gives us the opportunity to calibrate scaling relations that link spectroscopic or photometric observables, accessible with telescopes, to the derived mass of the galaxy clusters \citep{Okabe10, Allen11, Giodini13, Sereno15X, Bellagamba19, Lesci_counts, Seppi24, Lesci25}. Although the galaxy bias remains difficult to model properly, and it is often treated as a nuisance parameter, measuring cluster masses enables us to estimate the effective bias of the cluster sample, $b_\mathrm{eff}$, as a function of the cosmological model \citep[see, for instance,][]{Paech17, Lesci22, Romanello24}, and therefore to break its degeneracy with $\sigma_8$. \\
\indent Current Stage-III photometric surveys, such as the Hyper Suprime-Cam (HSC) Subaru Strategic Program \citep[HSC-SSP; see][]{Aihara18}, the Dark Energy Survey \citep[DES; see][]{DES16, Bechtol26} and the Kilo Degree Survey \citep[KiDS; see][]{deJong17, Kuijken19, Wright24}, provide deep and multi-band imaging over thousands of square degrees, and a large number of galaxies that can be used successfully for clustering, weak-lensing and, in general, cosmological analyses. These surveys also represent a valuable testing ground for models and methods that aim to maximise the amount of cosmological signal that can be extracted from ongoing Stage-IV surveys, such as the Vera C. Rubin Observatory Legacy Survey of Space and Time \citep[LSST; see][]{MandelbaumLSST, Ivezic19LSST} and the \textit{Euclid} mission \citep{Blanchard20, Scaramella22, Mellier25}. \\
\indent In this paper, we conduct a two-point correlation function analysis based on the fourth data release of the Kilo Degree Survey (hereafter referred to as KiDS-1000). The catalogue of galaxy clusters, identified by the Adaptive Matched Identifier of Clustered Objects \citep[AMICO; see][]{Bellagamba18} algorithm, has been described in \citet{Maturi25}, while the first cosmological results, derived from cluster abundance, as well as the weak-lensing calibration of the scaling relation, are presented in \citet{Lesci25}. In order to measure the appropriate statistical quantities and to run a full Bayesian inference, we make use of \textsc{CosmoBolognaLib} \citep{MarulliCBL}\footnote{\url{https://gitlab.com/federicomarulli/CosmoBolognaLib}}, a set of {\em free software} C++ and Python libraries.\\
\indent This work is part of a series of papers that employ AMICO KiDS-1000 clusters for both astrophysical and cosmological outcomes \citep[see, for example,][]{Maturi25, Lesci25, Lesci26}. It is organised as follows: in Sect. \ref{Sect_data} we present the AMICO KiDS-1000 cluster catalogue. In Sect. \ref{sect_twop} we describe the methodology adopted to measure and model the cluster two-point correlation function and its multipoles. In Sect. \ref{sect_cosmological_results} we discuss the results of the Bayesian inference, comparing them with publications from galaxy clusters cosmology available in the literature and with previous studies based on the KiDS survey. Finally, in Sect. \ref{Conclusions} we draw our conclusions.   
\section{Data: the AMICO KiDS-1000 galaxy cluster catalogue}
\label{Sect_data}
\begin{figure*}
   \centering
   \includegraphics[width=\hsize]{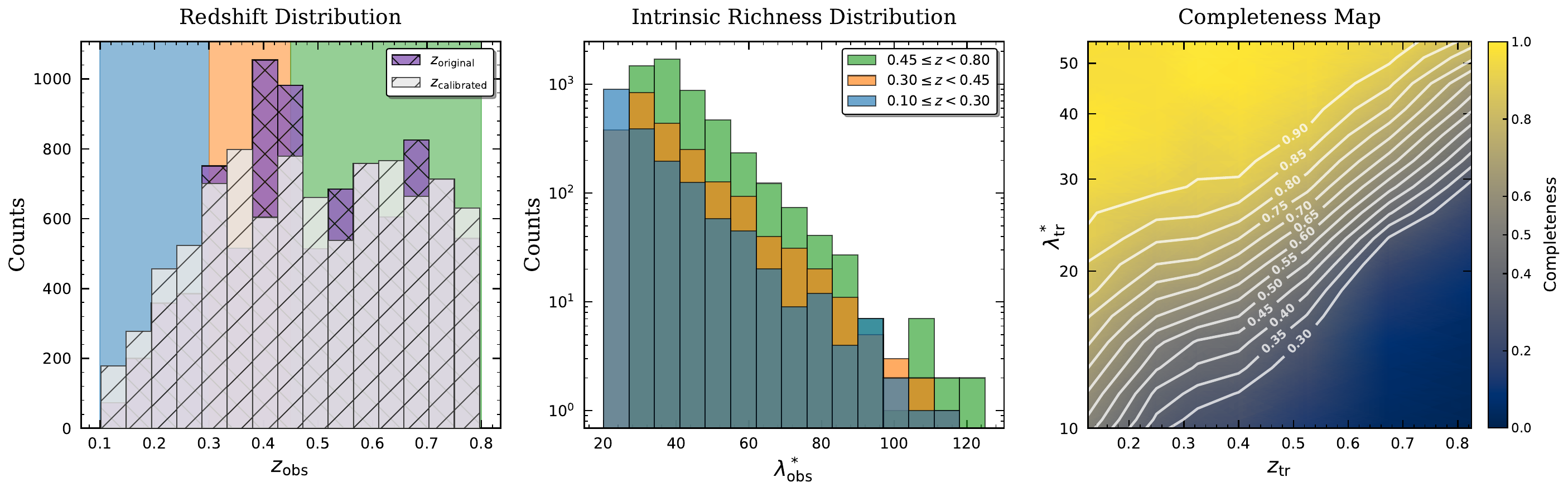}
      \caption{\textit{Left}: redshift distribution of the AMICO KiDS-1000 galaxy clusters. The purple histogram shows the original redshifts, before the spectroscopically based calibration was applied, as described in \citet{Maturi25}. The white overlaid histogram represents the bias-corrected photometric redshifts. The blue, orange and green  regions correspond to the three different redshift bins. \textit{Middle}: intrinsic richness distribution, after the $\lambda^*_\mathrm{obs}$ selection is applied to the photometric redshift bins. \textit{Right}: completeness map derived from mock catalogues, as a function of true redshift and intrinsic richness. The levels for the contours and their corresponding values are also shown in white.}
    \label{Fig_distributions}
   \end{figure*}
KiDS is a European Southern Observatory (ESO) optical survey specifically designed for cosmological applications. It collected observations in four photometric bands: \textit{u, g, r} and \textit{i}, using the OmegaCAM wide-field imager \citep{Kuijken11} mounted on the 2.6 m VLT Survey Telescope (VST), at the Paranal Observatory. In this paper, we focus on the latest cluster catalogue based on the fourth data release, named KiDS-1000 \citep{Kuijken19} that covers an area of 1006 deg$^2$, divided into two patches: one equatorial (KiDS-N) and the other across the South Galactic Pole (KiDS-S). KiDS-1000 represents a significant improvement compared to the previous release, KiDS-DR3 \citep{deJong17}. In fact, it also includes the photometry of the near-IR ($ZYJHK_s$) VIKING images \citep{Edge13}, from the ESO VISTA telescope. Moreover, the sky coverage is less fragmented and more homogeneous, and the area has increased by more than a factor of two.\\
\indent The limiting magnitudes ($5\sigma$ in
a 2" aperture) for the \textit{ugri} bands are $24.23\pm 0.12$, $25.12\pm 0.14$, $25.02\pm 0.13$ and $23.68\pm 0.27$, respectively. As in KiDS-DR3 \citep{Maturi19}, a magnitude cut is applied to reject galaxies with $r>24$, according to the $r$-band limit of the shallower observational tiles, in order to further ensure the homogeneity of the sample \citep{Maturi25}. \\
\indent AMICO \citep{Bellagamba18} is an optimal filtering algorithm for cluster detection, based on the convolution of the 3D galaxy distribution with a redshift-dependent filter. For the considered run, it relies only on the magnitude and coordinates of the galaxies, which are weighted according to their photometric redshift distribution, as provided for KiDS data set by the \texttt{Bayesian Photometric
Redshift} code \citep[BPZ,][]{Benitez2000}. Because of the ongoing cluster assembly process -- involving younger, bluer stellar populations and active star formation -- at high $z$ the cluster red sequence may not be clearly defined. Therefore, a colour-based selection is avoided, ruling out possible biases in the detection \citep{Adam19, Maturi19}. This makes the sample more comprehensive and less biased for cosmological studies. In addition to KiDS, AMICO has also been used to look for galaxy groups and clusters in the Javalambre-Physics of the Accelerating Universe Astrophysical Survey \citep[J-PAS,][]{Maturi23} and COSMOS fields \citep{Toni24, Toni25} and is one of the two cluster identification algorithms selected to process the photometric galaxy catalogues of the \textit{Euclid} mission \citep{Adam19, Bhargava26}, due to its well tested and excellent performance.\\
\indent During the processing of the KiDS-1000 data, regions affected by bright stars, reflections, and artefacts are masked, resulting in a final effective area of 838.78 deg$^2$. Clusters have been detected by sampling the sky in 0.3-arcmin steps and in redshift intervals of $\Delta_z = 0.01$, which corresponds to the redshift resolution of AMICO. Due to the long tails of the galaxy photometric redshift distribution, the search is carried out within $0.05<z<1.2$ and, to avoid edge effects, in overlapping tiles, with a 0.1-degree buffer zone on each side. The final catalogue contains 23965 detections with $S/N\geq3.5$, in the redshift range $0.1<z<0.9$. \\
\indent In this work, we use the intrinsic richness, $\lambda^*$, as a mass proxy. This quantity is provided as an output by AMICO, and it is defined as the sum of the membership probabilities:
\begin{equation}
\label{intrinsic_richness}
    \lambda^*_ j=\sum_{i=1}^{N_{gal}} P_i(j) \quad \textrm{with} \quad \begin{cases}
m_i<m^*(z_j)+1.5 \\
r_i(j)<R_{200}(z_j),\\
\end{cases}
\end{equation}
where $P_i(j)$ is the probability assigned by AMICO to the $i$-th galaxy of being part of a given detection, $j$. It is evaluated according to the explicit conditions on the magnitude, $m_i$, and on the distance of the $i$-th galaxy from the centre of the $j$-th cluster. In particular, $R_{200}(z_j)$ is the radius of the sphere in which the mean density is 200 times the critical density at redshift $z_j$, enclosing the characteristic mass of a galaxy cluster, $M_{200}=10^{14} h^{-1}$ M$_\odot$. The limit $m^*$ is a function of redshift and represents the typical $r$-band magnitude used in the Schechter function of the cluster model employed by the AMICO algorithm \citep[for more details, see][]{Maturi19}. Thus, for a given detection, the intrinsic richness roughly corresponds to the expected number of visible galaxies, under the explicit condition expressed by Eq. \eqref{intrinsic_richness}. The threshold $m^* + 1.5$ is significantly brighter than the magnitude limit of the survey and makes the intrinsic richness a redshift-independent quantity. However, as clearly assessed in \citet{Maturi25}, at $z > 0.8$ it approaches the magnitude limit of $r = 24$, resulting in a progressively unstable $\lambda^*$. For this reason, the conservative cut from the weak-lensing analysis conducted by \citet{Lesci25} is applied throughout the work. This restricts the catalogue to the range $0.1\leq z<0.8$, and adopts an additional selection in the mass proxy, to account for the loss in purity and completeness at high redshift, so that the final sample contains 9049 objects. Therefore, we split the catalogue in three redshift bins, $z\in[0.1, 0.3)$, $z\in[0.3, 0.45)$, $z\in[0.45, 0.8)$; in the corresponding ranges, only clusters with observed intrinsic richness $\lambda^*_\mathrm{obs}\geq 20,25,30$ are kept. With this choice the statistics is maximised, resulting in 1767, 2240 and 5042 candidates in each interval. \\
\indent The purity and the completeness are measured using the Selection Function extrActor code \citep[\textsc{SinFoniA},][]{Maturi25}, which produces a mock sample according to the galaxy probabilistic membership association, $P_i(j)$, and to the known properties of KiDS-1000, such as the galaxy density, photometry, photo-$z$ errors and the survey mask. The sky positions and redshifts in the mocks are only slightly and randomly perturbed before the AMICO algorithm is run. This means that the mock clusters preserve their auto-correlation and at the same time their cross-correlation with the unperturbed field galaxies which, by construction, retain their intrinsic clustering. Statistical properties of the purity and completeness are derived by comparing the original detections with the mock-based ones, as described in more detail in \citet{Lesci25} and \citet{Maturi25}. The completeness, $\mathcal{C}_\mathrm{clu}$, is interpolated from a grid of 8 redshift and 50 intrinsic richness bins. It is higher than $70\%$, $75\%$ and $35\%$, in the corresponding redshift intervals \citep{Lesci25, Maturi25}, while the purity is larger than $96\%$ in the whole cluster sample. \\
\indent When available, spectroscopic data from the Galaxy and Mass Assembly \citep[GAMA,][]{Driver11, Liske15} survey, augmented with higher redshift information from \citet{vandenBusch22}, were also used in the cluster redshift calibration, since they partially overlap with the KiDS-1000 footprint mask \citep{Bellstedt20}. In practice, GAMA redshifts are assigned to clusters if they are associated with at least one spectroscopic galaxy that has a photometric counterpart with an extremely high AMICO membership probability. This procedure enabled \citet{Maturi25} to correct the biases in the clusters photo-$z$, especially at $z<0.4$, and to quantify the photo-$z$ errors from the RMS of the bias-corrected photometric redshifts. \\
\indent In the left and middle panels of Fig. \ref{Fig_distributions} we show the KiDS-1000 cluster redshift and intrinsic richness distributions, after the $S/N$ and $\lambda^*_\mathrm{obs}$ selections are applied. In the right panel, we also report the completeness map derived from mock catalogues, as a function of the true redshift and intrinsic richness. After the calibrations, the conditional probability of having a photometric redshift, $z_\mathrm{obs}$, given a true redshift, $z_\mathrm{tr}$, is $P(z_{\mathrm{obs}}|z_\mathrm{tr})$, a Gaussian distribution, with mean $z_\mathrm{tr}$, and standard deviation equal to:
\begin{equation}
\label{eq_sigmaz}
    \sigma_z = 0.014(1 + z_\mathrm{tr}), 
\end{equation}
corresponding to $0.017$ in the first, $0.019$ in the second and $0.023$ in the third redshift bin, respectively; this represents an improvement with respect to the precision achieved in KiDS-DR3 photometry, $ \sigma_z = 0.02(1 + z_\mathrm{tr} )$. \\   
\section{The two-point correlation function of the AMICO KiDS-1000 cluster catalogue}
\label{sect_twop}
In this section, we describe the methods used to measure and model the two-point correlation function.

\label{two-point}
\subsection{Measure}
\label{Section_measure_twop}
\begin{figure*}
   \centering
   \includegraphics[width=\hsize]{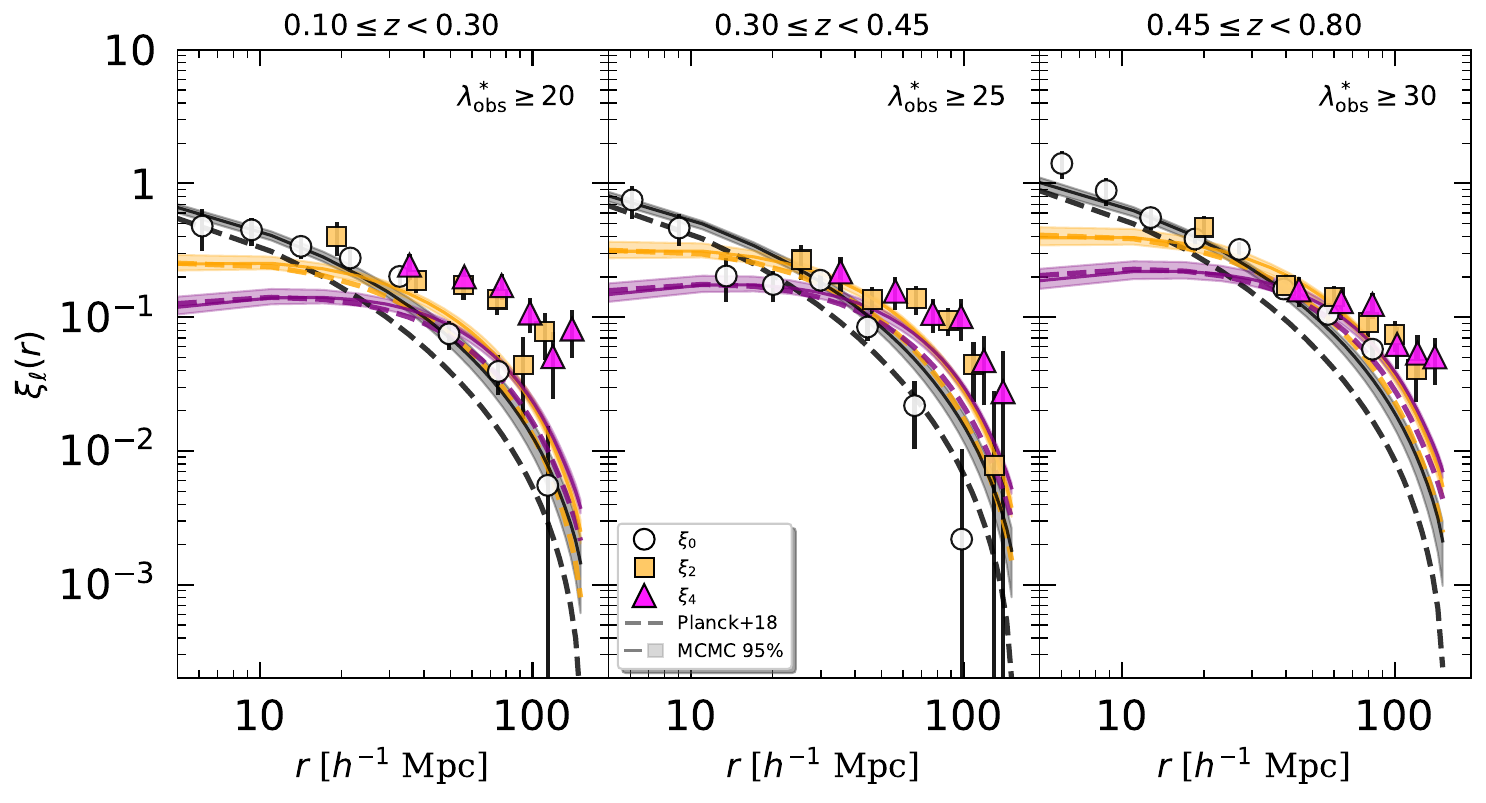}
      \caption{From left to right: multipoles of the two-point correlation function in the redshift bins $0.10\leq z<0.30$, $0.30\leq z<0.45$ and $0.45\leq z<0.80$, with $\lambda^*_\mathrm{obs}\geq 20,25,30$, respectively. The white circles represent the monopole, $\xi_0$, the orange squares the quadrupole, $\xi_2$, and the purple triangles the hexadecapole, $\xi_4$. The errors are obtained with the bootstrap method. For each $\xi_\ell$, the models are plotted with the corresponding colours. In particular, the dashed lines refer to the Planck+18 $\Lambda$CDM cosmology, the solid lines to the best fit, derived from the median of the combined posterior distribution, while the shaded regions highlight the 95\% confidence levels. }
    \label{Fig_2pcf}
   \end{figure*}
We can define the 3D two-point correlation function, $\xi(r)$, from the joint probability of finding two clusters in the comoving volumes $\mathrm{d} V_1$ and $\mathrm{d} V_2$:
\begin{equation}
    \mathrm{d} P_{12}(r)= n_{V}^2[1+\xi(r)]\mathrm{d} V_1 \mathrm{d} V_2,
\end{equation}
where $n_V$ is their mean comoving number density. It represents the excess probability, relative to that expected from a random distribution, to have a pair of galaxy clusters separated by a distance $r$. \\
\indent The two-point correlation function multipoles can be measured for $\ell=0-2-4$ using the \citet[LS,][]{Landy93} estimator:
\begin{equation}
    \hat{\xi_\ell}(r)=\frac{DD_\ell(r)+RR_\ell(r)-2DR_\ell(r)}{RR_0(r)},
\end{equation}
where 
$DD_\ell(r)$, $RR_\ell(r)$, and $DR_\ell(r)$ are the normalised data-data, random-random, and data-random pair counts in the radial bin $r \pm \Delta r/2$, respectively, weighted with the corresponding Legendre polynomials, $\mathcal{L}_{\ell}$. \\
\indent 
To limit shot noise effects, we build a random catalogue containing 70 times more clusters than the final AMICO KiDS-1000 catalogue. The angular coordinates are extracted within the survey observational tiles, using the same masks adopted in \citet{Maturi25}, for AMICO detections. To avoid generating an excess of random clusters, due to the overlap between tiles in Dec, we construct a high-resolution \textsc{Healpix} mask with $N_\mathrm{side}=4096$, assigning an equal weight of 1 to all the unmasked regions. In the $k$-th redshift bin, the Hartlap-corrected covariance matrix \citep{Hartlap07} is estimated through $N_\mathrm{BT}=500$ bootstrap resamplings, with replacement, considering a number of regions equal to the number of observational tiles, namely $N_\mathrm{reg}=1006$   \citep{Norberg09}:
\begin{equation}
    \label{eq_covariance}
    C_{(k)\ell\ell,\,ij}=\frac{1}{N_{\mathrm{BT}}-1}\sum_{m=1}^{N_{\mathrm{BT}}}(\hat{\xi}_{\ell,i}^{m, (k)}-\overline{\hat{\xi}}^{(k)}_{\ell, i})(\hat{\xi}_{\ell,j}^{m,(k)}-\overline{\hat{\xi}}^{(k)}_{\ell, j}),
\end{equation}
where the indices $i$ and $j$ mark the spatial bins involved in the measurement of the $m$-th realisation, with expectation values $\overline{\hat{\xi}}_{\ell, i}=\sum_{m=1}^{N_\mathrm{BT}} \hat{\xi}_{\ell,i}^{m}/N_{\mathrm{BT}}$. 
\\
\indent The monopole of the two-point correlation function is measured in 8 logarithmic bins ranging from $r_\mathrm{min}=5$ $h^{-1}$ Mpc to $r_\mathrm{max}=140-120-100$ $h^{-1}$ Mpc, from the lowest to the highest redshift. This choice places greater relevance on small separations, which contain most of the cosmological information, while at the baryon acoustic oscillations (BAO) scale are given less weight. As verified by \citet{Sereno15}, \citet{GarciaFarieta20} and \citet{Romanello25}, photo-$z$ errors introduce a scale-dependent damping, resulting in a change in the slope of the two-point correlation function and in a suppression of the BAO peak around $100$ $h^{-1}$ Mpc. The impact on the quadrupole and hexadecapole moments is even greater: the former changes sign and becomes positive, while the latter is no longer compatible with zero and takes on a dominant contribution. In other words, the photo-$z$ errors determine the transfer of the cosmological signal not only between different scales, but also between different multipoles.\\
\indent The discretisation of the redshift introduced by AMICO, $\Delta_z=0.01$, mentioned in Sect. \ref{Sect_data}, deserves a proper discussion. In fact, it produces a damping effect comparable to that of photo-$z$ errors, and it can be treated as a constant uncertainty equal to approximately half of the $\Delta_z$ -- in this case $0.005$ -- which is of minor significance when added in quadrature to $\sigma_z$. Moreover, since along the line of sight the galaxy cluster pairs are counted only in multiples of the discretisation step, the correlation function is subject to several oscillations, becoming unstable, especially at small scales and for higher multipoles. This requires cutting the radial interval at progressively larger $r_\mathrm{min}$. In particular, for the quadrupole we have chosen $r_\mathrm{min}=[10, 15, 10]$ $h^{-1}$ Mpc and $r_\mathrm{max}=[120, 140, 130]$ $h^{-1}$ Mpc, while for the hexadecapole $r_\mathrm{min}=[25, 25, 35]$ $h^{-1}$ Mpc and $r_\mathrm{max}=[150, 150, 150]$ $h^{-1}$ Mpc. The upper limit was set trying to prioritise large scales, ensuring at the same time the stability of the measurements. \\
\indent In Fig. \ref{Fig_2pcf}, we show the first three even multipoles of the two-point correlation function, for every redshift bin. They are compared with the model presented in Sect. \ref{Sect_twop_model}, computed assuming the cosmology provided by \citet{Planck20}, Table 2, TT, TE, EE + lowE + lensing (hereafter Planck+18), and with the result of our Bayesian analysis of Sect. \ref{sect_cosmological_results}.
\subsection{The two-point correlation function model}
\label{Sect_twop_model}
In this section, we describe the theoretical models used to derive cosmological constraints from the clustering signal of galaxy clusters. The multipole moments of the two-point correlation function are modelled as follows:
\begin{equation}
    \xi_{\ell}(r)=\frac{i^{\ell}}{2 \pi^2} \int \mathrm{d} k \, k^2 P_{\ell}(k) j_{\ell}(k r),
\end{equation}
where $k$ is the comoving wavevector, $j_\ell$ are the spherical Bessel functions and $P_\ell(k)$ is the $\ell$-th multipole of the power spectrum, defined by: 
\begin{equation}
    \label{eq_pkl}
    P_{\ell}(k)=\frac{2 \ell+1}{2\alpha^2_\perp \alpha_\parallel} \int ^{+1} _{-1}\mathrm{d} \mu P\left(k', \mu'\right) \mathcal{L}_{\ell}\left(\mu\right). 
\end{equation}
Here, the Legendre polynomials are computed at $\mu$, the cosine of the angle between the line of sight and the $k$-vector. The anisotropic power spectrum, $P(k', \mu')$, is corrected for geometric distortions through the coefficients $\alpha_\perp$ and $\alpha_\parallel$, with values:
\begin{equation} 
    \begin{pmatrix}
        \alpha_{\perp} \\[4pt]
        \alpha_{\|}
    \end{pmatrix}
    =
    \begin{pmatrix}
        \dfrac{D_A(z)}{D_A^{\mathrm{fid}}(z)} \\[10pt]
        \dfrac{\mathrm{H}^{\mathrm{fid}}(z)}{\mathrm{H}(z)}
    \end{pmatrix}, 
\end{equation}
where $D_A(z)$ is the angular diameter distance, $H(z)$ is the Hubble parameter, while quantities with the superscript `fid' are computed at the fiducial cosmology, taken from Planck+18. Indeed, measurements of the two-point correlation function need a cosmological assumption in order to convert the observed redshifts and positions into three-dimensional comoving distances. However, this fiducial cosmology may be biased with respect to the underlying, unknown one; therefore, the computation of the power spectrum requires a rescaling \citep{Beutler14}. In particular: 
\begin{equation}
    k^{\prime}=\frac{k}{\alpha_{\perp}}\left[1+\mu^2\left(\frac{\alpha_{\perp}^2}{\alpha_{\|}^2}-1\right)\right]^{1 / 2},
\end{equation}
and: 
\begin{equation}
    \mu^{\prime}=\mu \frac{\alpha_{\perp}}{\alpha_{\|}}\left[1+\mu^2\left(\frac{\alpha_{\perp}^2}{\alpha_{\|}^2}-1\right)\right]^{-1 / 2}.
\end{equation}
\indent Equation \eqref{eq_pkl} also accounts for RSDs, associated with the peculiar velocities of large-scale structures, and photo-$z$ errors. The anisotropic power spectrum is defined starting from the dispersion method \citep{Peacock94, Pezzotta17, Tsaprazi26} as follows: 
\begin{equation}
    P(k, \mu)=D\left(k, \mu, f, \sigma_\mathrm{tot}\right)\left(1+\frac{f}{b_\mathrm{eff}} \mu^2\right)^2 b_\mathrm{eff}^2 P_\mathrm{lin}(k),
\end{equation}
in which the Kaiser factor in parentheses, that depends on the linear growth rate, $f$, and on the effective bias, $b_\mathrm{eff}$ (see Sect. \ref{Effective_bias}), describes the coherent infall of structures towards collapsing systems; $P_\mathrm{lin}(k)$ is the dark matter linear power spectrum computed through the parametrisation presented in \citet{Eisenstein98}. After the Fourier transform, below the BAO scale, the resulting two-point correlation function differs from the one estimated with the Boltzmann solver CAMB \citep{Lewis00} by up to a maximum of 2.7, 3.8 and 3.1\%, for $\ell = 0-2-4$, respectively. The damping factor, $D$, is modelled by $\sigma_\mathrm{tot}(z)$:
\begin{equation}
    D\left(k, \mu, f, \sigma_\mathrm{tot}\right)=\mathrm{exp}(-k^2\mu^2\sigma_\mathrm{tot}^2),
\end{equation}
a combination of two different contributions:
\begin{equation}
\label{eq_total_error}
    \sigma_\mathrm{tot}^2=f^2 \sigma_\mathrm{v}^2+\sigma^2(z).
\end{equation}
\indent The former,  which includes the linear growth rate, $f$, depends on the dynamics of large-scale structures, while the latter on the specifications of the survey and the detection algorithm. By analysing, from a physical point of view, the total uncertainty that contributes to the damping, we notice that the former term in Eq. \eqref{eq_total_error} quantifies the FoG effect, subdominant since galaxy clusters are only weakly affected by nonlinear dynamics at the scale considered, and it is based on a linear-order estimate of
the one-dimensional velocity dispersion \citep{Taruya10}:
\begin{equation}
    \sigma_{\mathrm{v}}^2=\frac{1}{3} \int \frac{\mathrm{d}^3 \boldsymbol{k}}{(2 \pi)^3} \frac{P_{\mathrm{lin}}(k, z)}{k^2}. 
\end{equation}
\indent It only contributes to the total error by $7\%$, in all the redshift bins. As it is negligible compared to the damping due to the photo-$z$ errors, we verified that it can be safely neglected \citep{Sereno15}. \\
\indent The second term of Eq. \eqref{eq_total_error} includes the discretisation of the observed redshift induced by AMICO, $\sigma_\Delta$, and the photo-$z$ damping, and can be written as: 
\begin{equation}
    \sigma(z)=\frac{c\sqrt{\sigma^2_z+\sigma_\Delta^2}}{H(z)},
\end{equation}
where $c$ is the speed of light and $\sigma_z$ is given by Eq. \eqref{eq_sigmaz}. Here, the contribution of $\sigma_\Delta$ is only $29\%$, $26\%$, and $22\%$ of the photo-$z$ uncertainty, in the three redshift bins. Although this effect is also subdominant, it is worth taking it into account in light of future improvements in the photometric accuracy of current and upcoming surveys, which will lead to a reduction in the expected value of $\sigma_z$.
\subsection{Effective bias of the cluster sample}
\label{Effective_bias}
Galaxy clusters are biased tracers of the underlying density distribution. This means that on linear scales, in the first approximation, their density field differs from the dark matter density field only by a multiplicative constant bias factor. \\
\indent The effective bias of the cluster sample links the parametrisation of the mass and redshift dependence of the bias, calibrated on simulations \citep[see, for example,][]{Tinker10, Castro24}, to the actual high-mass tail in the distribution of dark matter haloes. It therefore depends on the halo mass function. The mass of a galaxy cluster is, in turn, statistically related to a photometrically identified mass proxy, bridging theoretical calibrations with the observed population. In particular, the intrinsic richness, $\lambda^*$, represents a natural choice for optically-selected clusters. In this paper, we make use of the mass-richness scaling relation calibrated by \citet{Lesci25}, through weak lensing and cluster number counts:
\begin{equation}
    P\left(\lambda_{\mathrm{tr}}^* \mid M, z_\mathrm{tr}\right)=\frac{1}{\ln (10) \lambda_{\mathrm{tr}}^* \sqrt{2 \pi} \sigma_{\mathrm{intr}}} \exp \left(-\frac{\left[\log \lambda_{\mathrm{tr}}^*-\mu\left(M, z_\mathrm{tr}\right)\right]^2}{2 \sigma_{\mathrm{intr}}^2}\right).
\end{equation}
\indent It is a log-normal probability density distribution, with constant standard deviation, $\sigma_\mathrm{intr}$. The mean, $\mu(M,z_\mathrm{tr})$, is expressed as follows: 
\begin{equation}
    \mu\left(M, z_\mathrm{tr}\right)=\alpha+\beta \log \frac{M}{M_{\mathrm{piv}}}+\gamma \log \frac{H\left(z_\mathrm{tr}\right)}{H\left(z_{\mathrm{piv}}\right)}+\log \lambda_{\mathrm{piv}}^*.
\end{equation}
\indent The mass-richness scaling relation consists of a normalisation constant $\alpha$, a slope $\beta$, and a redshift evolution $\gamma$. The pivotal quantities reflect the typical central values of their respective distributions and are equal to $M_\mathrm{piv}=10^{14}$ $h^{-1}$ M$_\odot$, $\lambda^*_\mathrm{piv}=50$ and $z_\mathrm{piv}=0.4$. Higher values of $\alpha$ and $\sigma_\mathrm{intr}$ tend to increase the effective bias of the cluster sample; vice versa, it decreases with the parameter $\beta$. The bias is also positively correlated with $\gamma$, but this dependence softens with redshift, being stronger in the bin $0.10\leq z<0.30$.\\
\indent Inside a given photometric redshift bin, $\Delta z_\mathrm{obs}$, we assume that the effective bias is constant and does not evolve:
\begin{equation}
    b_\mathrm{eff}\left(\Delta z_\mathrm{obs}\right)=
    \frac{
      \begin{aligned}
      &\int_0^{\infty}\!\mathrm{d} z_\mathrm{tr} \frac{\mathrm{d^2}V}{\mathrm{d} z_\mathrm{tr}\mathrm{d} \Omega}
       \int_0^{\infty}\!\mathrm{d} M
       n(M, z_\mathrm{tr})
       b(M,z_\mathrm{tr})
       \mathcal{K}\!\left(M, z_\mathrm{tr}\right)
      \end{aligned}
    }{
      \begin{aligned}
      &\int_0^{\infty}\!\mathrm{d} z_\mathrm{tr} \frac{\mathrm{d^2}V}{\mathrm{d} z_\mathrm{tr}\mathrm{d} \Omega} 
       \int_0^{\infty}\!\mathrm{d} M
       n(M, z_\mathrm{tr})
       \mathcal{K}\!\left(M, z_\mathrm{tr}\right)
      \end{aligned}
    },
\end{equation}
where $n(M,z)$ is the halo mass function, according to the parametrisation provided by \citet{Tinker08}, $b(M,z)$ is the halo bias computed using the fitting formulae of \citet{Tinker10}. Finally, the integration kernel, $\mathcal{K}(M,z_\mathrm{tr})$, takes into account the completeness of the cluster sample, the convolutions with the mass-richness scaling relation and the dependencies of the observed quantities on the true quantities, as evaluated using mocks:  
\begin{equation}
\label{eq_kernel}
\begin{gathered}
    \mathcal{K}\!\left(M, z_\mathrm{tr}\right) = 
    \int_0^{\infty}\!\mathrm{d} \lambda_{\mathrm{tr}}^*\,
    \mathcal{C}_\mathrm{clu}\!\left(\lambda_{\mathrm{tr}}^*, z_\mathrm{tr}\right)
    P\!\left(\lambda_{\mathrm{tr}}^* \mid M, z_\mathrm{tr}\right)  \\
    \times \int_{\Delta \lambda^*_\mathrm{obs}}\!\mathrm{d} \lambda_{\mathrm{obs}}^*\,
    P\!\left(\lambda_{\mathrm{obs}}^* \mid \lambda_{\mathrm{tr}}^*, z_\mathrm{tr}\right) \\
    \times \int_{\Delta z_\mathrm{obs}}\!\mathrm{d} z_\mathrm{obs}\,
    P\!\left(z_\mathrm{obs} \mid z_\mathrm{tr}\right).
    \end{gathered}
\end{equation}
\indent We remind that $P(z_\mathrm{obs}|z_\mathrm{tr})$ is the conditional probability density of having a cluster in a neighbourhood of the observed redshift $z_\mathrm{obs}$ given the true one, $z_\mathrm{tr}$. It is a Gaussian probability distribution with a mean corresponding to $z_\mathrm{tr}$ and a standard deviation given by $\sigma_z$, as in Eq. \eqref{eq_sigmaz}. By combining the discretisation uncertainties in quadrature, that is $\sqrt{\sigma_\Delta^2+\sigma_z^2}$, we verify that the effective bias remains unchanged. Conversely, $P(\lambda^*_\mathrm{obs}|\lambda^*_\mathrm{tr}, z_\mathrm{tr})$ in Eq. \eqref{eq_kernel} is a normal distribution, whose mean depends on the redshift and intrinsic richness, whereas the standard deviation is only a function of $\lambda^*_\mathrm{tr}$: 
\begin{equation}
\begin{split}
    \mu_{\lambda^*} &= \lambda_{\mathrm{tr}}^* + \lambda_{\mathrm{tr}}^* \exp \left[-\lambda_{\mathrm{tr}}^*\left(\bar{A}_\mu+\bar{B}_\mu z_{\mathrm{tr}}\right)\right], \\
    \sigma_{\lambda^*} &= \bar{A}_\sigma \lambda_{\mathrm{tr}}^* \exp \left(-\bar{B}_\sigma \lambda_{\mathrm{tr}}^*\right).
\end{split}
\end{equation}
The parameters $\bar{A}_\mu=0.198$, $\bar{B}_\mu=-0.179$, $\bar{A}_\sigma=0.320$, $\bar{B}_\sigma=0.011$ were also derived by \citet{Lesci25}. \\
\indent The completeness, $\mathcal{C}_\mathrm{clu}$, already presented in Fig. \ref{Fig_distributions}, is the number of cluster detections with a true counterpart divided by the number of true clusters \citep{Lesci25, Maturi25}, evaluated for different redshift and richness bins from the data-driven mock cluster catalogue produced with \textsc{SinFoniA} \citep{Maturi25}. This factor proportionally penalises systems with low intrinsic richness, which consequently have a smaller mass. As a result of suppressing the low-mass end of $n(M,z)$, the halo bias is weighted with a population shifted towards higher masses, resulting in an increase of $b_\mathrm{eff}$. Since both the completeness of detection and the richness tend to be lower at high $z$, the ratio of the bias estimated at the Planck+18 cosmology for the complete population to that for the incomplete population decreases, going from $0.96$ for the first two bins to $0.93$ in the range $z\in[0.45, 0.8)$. \\
\section{Cosmological analysis}
\label{sect_cosmological_results}
\begin{table*}[h!]
\caption{Parameters, prior and posterior median and percentiles of the cosmological analyses.}   
\label{table_par}      
\centering          
\begin{tabular}{c c c c}     
\hline\hline       
Parameter & Description & Prior & Posterior \\ 
\hline                    
    $\Omega_\mathrm{CDM}$  & Cold dark matter density parameter & $[0.1, 0.5]$ & $0.19^{+0.02}_{-0.02}$\\
    
    $\sigma_8$  & Amplitude of the mass variance on the scale of 8 $h^{-1}$ Mpc & $[0.5, 1.5]$ & $0.86^{+0.13}_{-0.10}$\\
    
    $\Omega_\mathrm{b}$  & Baryon density parameter & $\mathcal{G}(0.0493, 0.0016)$ & -\\

    $n_\mathrm{s}$  & Primordial spectral index & $\mathcal{G}(0.965, 0.021)$ & -\\

    $h\equiv H_0/(100 \, \mathrm{km/s/Mpc})$  & Normalised Hubble constant & $\mathcal{G}(0.7, 0.1)$ & -\\
    
    $\alpha$  & Normalisation of the mass-richness scaling relation & $\mathcal{G}(-0.17, 0.02)$ & -\\

    $\beta$  & Slope of the mass-richness scaling relation & $\mathcal{G}(0.55, 0.02)$ & -\\

    $\gamma$  & Redshift evolution of the mass-richness scaling relation & $\mathcal{G}(0.32, 0.27)$ & -\\

    $\sigma_{\mathrm{intr}}$  & Intrinsic scatter of the mass-richness scaling relation & $\mathcal{G}(0.052, 0.019)$ & -\\

    $\Omega_\mathrm{m}$  & Total matter density parameter & - & $0.24^{+0.02}_{-0.02}$\\
    
    $S_8\equiv \sigma_8(\Omega_\mathrm{m}/0.3)^{0.5}$& Structure growth parameter & - & $0.77^{+0.10}_{-0.08}$\\
\hline                  
\end{tabular}
\tablefoot{In the first and second columns, we present the list of the cosmological parameters, with the corresponding definitions, respectively. In the third column, the square brackets delimit the intervals of the uniform priors, while the letter $\mathcal{G}$ represents a normal distribution, with its mean and standard deviation explicitly expressed. Finally, the 1D posterior distributions are indicated, with their $1\sigma$ uncertainty level. }
\end{table*}
In Table \ref{table_par}, we list the complete set of parameters used in the cosmological analysis. Our aim is to derive constraints on the cold dark matter density, $\Omega_\mathrm{CDM}$, and $\sigma_8$, for which the two-point correlation function constitutes a highly sensitive probe. For this reason, we adopt non-informative priors on them. The total matter density parameter is then derived as $\Omega_\mathrm{m}=\Omega_\mathrm{CDM}+\Omega_\mathrm{b}$, and discussed throughout this section. For the normalised Hubble constant, $h$, we assume a Gaussian prior centred around 0.7, with a standard deviation of 0.1, wide enough to be agnostic with regard to the Hubble tension \citep[][for a review]{DiValentino21}. For the remaining parameters of the $\Lambda$CDM model, that are the baryon density, $\Omega_\mathrm{b}$, and the primordial spectral index, $n_\mathrm{s}$, we choose Gaussian priors around the median values from Planck+18, with standard deviations comprehensive to 5$\sigma$. The same choice is made for the normalisation, the mass and redshift evolution, and the intrinsic scatter of the scaling relation: $\alpha$, $\beta$, $\gamma$, $\sigma_\mathrm{intr}$, respectively, whose means and $1\sigma$ standard deviations were derived from the posterior distributions of the joint analysis of weak lensing and cluster abundances, as presented in \citet{Lesci25}.\\
\indent Similarly, our goal is to investigate the derived parameter, $S_8\equiv\sigma_8(\Omega_\mathrm{m}/0.3)^{0.5}$, in the light of the statistically significant discrepancy \citep[][for reviews]{Cosmoverse25, Pantos26} that has emerged between the cosmological results by Planck and other joint CMB analyses \citep[see, for example, the combination of TT, EE and TE power spectra from Planck 2018, ACT Data Release 6 and SPT-3G, presented in][]{Abbott26}, and some Stage-III cosmological results, from cosmic shear \citep{Asgari21} or $3\times2$pt statistics \citep{Abbott22}, which has recently been alleviated by KiDS-Legacy cosmic shear \citep{Wright25}. \\
\indent The Bayesian approach employed here is based on the implicit assumption of independent redshift bins. We verified that this choice is supported by a cross-covariance analysis, so that the final likelihood is the product of the individual, $k$-th, Gaussian likelihoods: 
\begin{equation}
\mathcal{L}_k\propto \exp(-\chi^2_k/2),
\end{equation}
where: 
\begin{equation}
    \chi^2_k=\sum_{\ell} \sum_{i=1}^{N_\ell} \sum_{j=1}^{N_{\ell}} 
    \left(\hat{\xi}_{\ell,i}^{(k)}-\xi_{\ell,i}^{(k)}\right) 
    (C^{-1}_{(k)})_{\ell \ell,\,ij} 
    \left(\hat{\xi}_{\ell,j}^{(k)}-\xi_{\ell,j}^{(k)}\right),
\end{equation}
being $N_\ell$ the number of spatial bins for the corresponding multipole, and $(C_{(k)})^{-1}_{\ell\ell,\,ij}$ the inverse of the covariance matrix for the $k$-th redshift bin, estimated via $N_{\mathrm{BT}}$ bootstrap resamplings of the observational tiles, according to Eq.~\eqref{eq_covariance}.\\
\begin{figure}[htbp]
\centering
\includegraphics[width=0.5\textwidth]{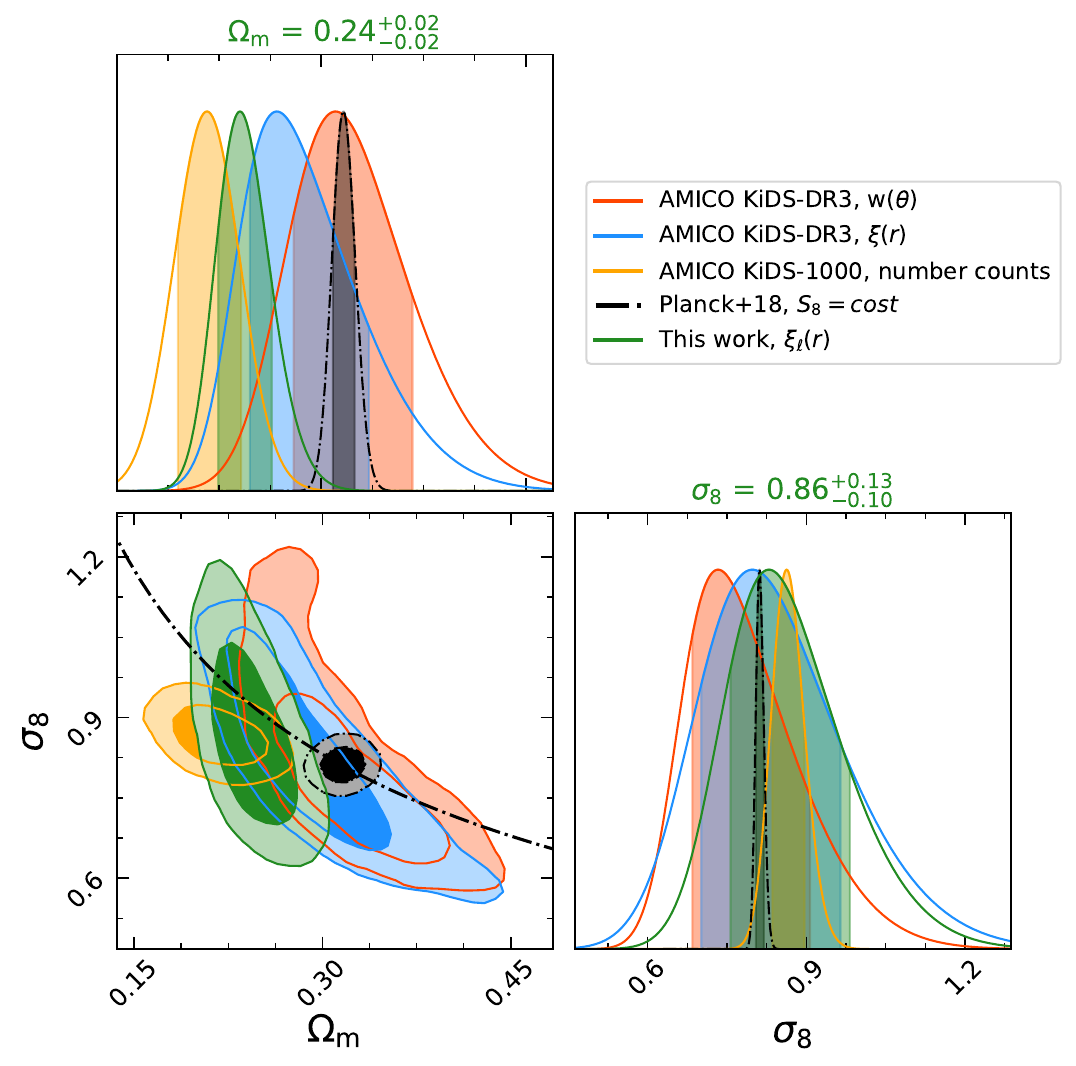}
\caption{Cosmological constraints in the $\Omega_\mathrm{m}-\sigma_8$ plane, with $68\%$ and $95\%$ confidence intervals, from 3D two-point correlation function multipoles (green contour). This result is compared to the recent findings from the same AMICO KiDS-1000 cluster catalogue, with number counts \citep[gold contour,][]{Lesci25}, from the previous 3D and angular clustering analyses in AMICO KiDS-DR3 \citep[pale blue and dark orange contours,][respectively]{Lesci22, Romanello24} and with CMB results from Planck+18 (black contour). For this last cosmology, the dot-dashed curve indicates points where the parameter $S_8$ is constant. In the 1D marginalised posterior distributions, the 68\% uncertainty levels are shaded.}
\label{Fig_corner}
\end{figure}
\indent From the Markov Chain Monte Carlo (MCMC) cosmological analysis of the two-point correlation function multipoles we find:
\begin{equation}
\begin{split}
\Omega_\mathrm{m}=0.24^{+0.02}_{-0.02}\,,\\
\sigma_8=0.86^{+0.13}_{-0.10}\,, 
\end{split}
\end{equation}
where the quoted values correspond to the medians, the $16$th and $84$th percentiles of the marginalised 1D distributions. The results in the $\Omega_\mathrm{m}-\sigma_8$ plane are shown in Fig. \ref{Fig_corner}, which also includes a comparison with the closest probes, such as the posteriors derived through 3D and angular clustering in AMICO KiDS-DR3 \citep{Lesci22, Romanello24} and the cluster number counts in AMICO KiDS-1000 \citep{Lesci25}. For the parameter $\sigma_8$, our results are in good agreement with previous constraints from the AMICO KiDS-DR3 clustering: $0.77^{+0.13}_{-0.09}$ for the angular - $w(\theta)$ - and $0.82^{+0.14}_{-0.12}$ for the 3D correlation function, respectively. However, the relative uncertainty is generally lower compared to them. In fact, it drops from a maximum of $17\%$ to around $15\%$. The same consistency is also found for the parameter $\Omega_\mathrm{m}$, for the 3D case, whereas it reduces to $2\sigma$ for $w(\theta)$. In both cases, the absolute uncertainty is halved, demonstrating a marked improvement in the constraining power of the correlation function. This can be quantified by introducing the Figure of Merit (FoM):
\begin{equation}
    \text{FoM} = \frac{1}{\sqrt{\det(C(\Omega_\mathrm{m},\sigma_8))}},
\end{equation}
defined in the plane $\Omega_\mathrm{m}-\sigma_8$ from the determinant of the covariance between the two parameters. It passes from the values of 297 and 298, measured from the 3D and 2D clustering analyses in AMICO KiDS-DR3 respectively, to the current value of 538, with a boost corresponding to a factor of approximately $1.8$.\\
\indent In our MCMC analysis, we treat the effective bias as a derived parameter, individually estimating it for each redshift bin. In particular, we get: $b_\mathrm{eff, 1} = 2.40^{+0.38}_{-0.36}$, $b_\mathrm{eff,2}=2.89^{+0.49}_{-0.46}$, $b_\mathrm{eff, 3}=3.68^{+0.64}_{-0.60}$. These values confirm the expected trend of growth with redshift, with a relative uncertainty associated with the bias that slightly rises, ranging from $15\%$ to $17\%$. In all three cases, the bias exhibits a robust correlation with $\Omega_\mathrm{m}$ and an even stronger anti-correlation with $\sigma_8$. Its dependence on the parameters of the scaling relation is also investigated. In fact, as previously mentioned in Sect. \ref{Effective_bias}, the effective bias increases moderately with $\alpha$ and $\sigma_\mathrm{intr}$, decreases mildly with $\beta$, while its correlation with $\gamma$ is generally weak and becomes less relevant with $z$. \\
\indent Our constraints are in excellent agreement with $\Omega_\mathrm{m}=0.24^{+0.03}_{-0.04}$ and $\sigma_8=0.86^{+0.07}_{-0.07}$, presented in \citet{Lesci_counts}, and with $\Omega_\mathrm{m}=0.22^{+0.02}_{-0.02}$ and $\sigma_8=0.86^{+0.03}_{-0.03}$, derived from the same AMICO KiDS-1000 cluster catalogue \citep{Lesci25}. Both of them are inferred with number counts and tend to favour a low value for $\Omega_\mathrm{m}$, in line with some Stage-III late-time probes \citep{Chiu23, Xiangchong23, Abbott25}, while comparison with Planck+18 reveals a $2\sigma$ tension in the matter density parameter. Although clustering is not as effective as number counts in determining the value of $\sigma_8$, the different orientation of their corner plots, as can be seen in Fig. \ref{Fig_corner}, reveals the full potential of probe combination \citep{Sartoris16, Fumagalli26, Martsen26}.\\
\indent Although the uncertainty in the individual parameters has decreased, it should be noted that we do not find any narrowing in the structure growth parameter marginalised posterior distribution, for which we get: 
\begin{equation}
    S_8=0.77^{+0.10}_{-0.08}.
\end{equation}
In other words, along the direction of degeneracy selected by $S_8$, we do not see any improvement compared to the values of $0.80^{+0.08}_{-0.08}$ and $0.80^{+0.08}_{-0.06}$ found in the AMICO KiDS-DR3 3D and angular two-point correlation functions, respectively \citep{Lesci22, Romanello24}. Again, this result is in line with the findings of \citet{Lesci25} where, despite significant reductions in statistical uncertainties and in the $\Omega_\mathrm{m}-\sigma_8$ FoM, the errors on $S_8$ are more stable, decreasing by only $0.01$. All the relevant comparisons between the current work and previous publications based on AMICO KiDS cluster catalogues are summarised in Table \ref{table:comparison}.\\
\begin{table*}[h!]
\caption{Cosmological constraints from AMICO KiDS cluster catalogues analyses, for $\Omega_\mathrm{m}$, $\sigma_8$, and $S_8$.}
\label{table:comparison}
\centering
\renewcommand{\arraystretch}{1.4}
\begin{tabular}{l c c c r}
\hline\hline
Probe & $\Omega_\mathrm{m}$ & $\sigma_8$ & $S_8$ & Reference \\
\hline
This work, $\xi_\ell(r)$              & $0.24^{+0.02}_{-0.02}$  & $0.86^{+0.13}_{-0.10}$ & $0.77^{+0.10}_{-0.08}$ & --  \\
AMICO KiDS-1000, number counts        & $0.22^{+0.02}_{-0.02}$  & $0.86^{+0.03}_{-0.03}$ & $0.74^{+0.03}_{-0.03}$ & \citet{Lesci25}       \\
AMICO KiDS-DR3, number counts         & $0.24^{+0.03}_{-0.04}$  & $0.86^{+0.07}_{-0.07}$ & $0.78^{+0.04}_{-0.04}$ & \citet{Lesci_counts}  \\
AMICO KiDS-DR3, $\xi(r)$             & $0.28^{+0.05}_{-0.04}$  & $0.82^{+0.14}_{-0.12}$ & $0.80^{+0.08}_{-0.08}$ & \citet{Lesci22}       \\
AMICO KiDS-DR3, $w(\theta)$          & $0.32^{+0.05}_{-0.04}$  & $0.77^{+0.13}_{-0.09}$ & $0.80^{+0.08}_{-0.06}$ & \citet{Romanello24}   \\
\hline
\end{tabular}
\tablefoot{1D marginalised distributions for $\Omega_\mathrm{m}$, $\sigma_8$, and $S_8 \equiv \sigma_8(\Omega_\mathrm{m}/0.3)^{0.5}$ from different AMICO KiDS cluster cosmological probes. All values are quoted as the median with 68\% confidence intervals. In \citet{Lesci25} and \citet{Lesci_counts} constraints are derived from cluster number counts and stacked weak-lensing profiles, on AMICO KiDS-1000 and AMICO KiDS-DR3, respectively. In \citet{Lesci22} and \citet{Romanello24} the 3D two-point and 2D angular correlation functions are used, both measured from the AMICO KiDS-DR3 catalogue.}
\end{table*}
\indent To confirm the robustness of our results, we repeat the analysis focusing on individual redshift bins. We find $\Omega_\mathrm{m}=0.23^{+0.03}_{-0.02}-0.29^{+0.04}_{-0.03}-0.22^{+0.03}_{-0.02}$, $\sigma_8=0.79^{+0.21}_{-0.14}-0.89^{+0.32}_{-0.20}-0.94^{+0.20}_{-0.15}$, within $0.10\leq z<0.30$, $0.30\leq z<0.45$ and $0.45\leq z<0.80$, respectively. Despite the broader uncertainty arising from the lower statistics involved, every result is consistent within 2$\sigma$. Finally, we obtain: $S_8=0.68^{+0.15}_{-0.10}-0.87^{+0.26}_{-0.15}-0.81^{+0.13}_{-0.11}$. As we can see, the first and third bins yield estimates of the matter density that are in excellent accordance, while the central range pushes it upwards. On the other hand, the first bin reduces $\sigma_8$, while most of the constraining power comes from the third bin, with a value for $S_8$ closer to the one given by the final combination. The same level of agreement is found when considering the redshift range $0.10\leq z < 0.45$ as a unique interval, with a more stringent intrinsic richness selection, $\lambda^*_\mathrm{obs}\geq30$.\\
\indent We also verify that our results are stable with respect to a different parametrisation of the halo mass function. Indeed, using the fitting formulae presented in \citet{Despali16}, the model computed with the Planck+18 cosmology differs no more than 2.7\%, 4.0\% and 3.3\%, for $\ell=0-2-4$, respectively. \\ 
\indent In Fig. \ref{Fig_comparison_S8} 
we compare our main outcome in the light of recent cosmological findings, both in the field of galaxy cluster cosmology, and more generally with other early- and late-Universe probes, with particular reference to the $S_8$ tension. From this point of view, the $S_8$ value derived from the Bayesian inference of the $\xi_\ell(r)$ AMICO KiDS-1000 shares a $1\sigma$ agreement with both the CMB based cosmological constraints, from Wilkinson Microwave Anisotropy Probe \citep[WMAP,][Table 3, WMAP-only, Year 9]{WMAP13} and Planck+18, as well as with the most recent findings based on cosmic shear in KiDS-1000 \citep{Asgari21} and KiDS-Legacy \citep{Wright25}. The same level of agreement is kept with respect to several works on cluster counts cosmology. For example, in the X-ray band, with the eROSITA first All-Sky Survey  \citep[eRASS1,][]{Ghirardini24} and eROSITA Final Equatorial Depth Survey \citep[eFEDS,][]{Chiu23}, in millimetric band with the combined South Pole Telescope cluster sample \citep[SPT,][]{Bocquet24}, which covers an effective area of 5200 deg$^2$. The abundance of optically-selected clusters, identified with the red-sequence Matched-filter Probabilistic Percolation algorithm \citep[redMaPPer,][]{Rykoff14, Rykoff16}, also produces very similar results, in the context of DES \citep{Abbott25} and, jointly with weak lensing and projected clustering, in HSC surveys \citep{Sunayama24}.\\
\indent Most of these studies focus on cluster abundances, as this is currently the most constraining approach. In fact, due to the broad uncertainty, 3D clustering alone is not yet able to provide competitive constraints with respect to the $S_8$ tension, despite the fact that the same AMICO KiDS-1000 data set has been able to reveal a $2.8\sigma$ discrepancy with respect to Planck+18 \citep{Lesci25}. Nevertheless, our work shows that the two-point correlation function provides valuable and independent confirmation. In the future, we plan to exploit the higher statistics ensured by the KiDS-Legacy \citep{Wright24} and \textit{Euclid} \citep{Sartoris16, Mellier25} sky coverage and sample size, the further reduction in photo-$z$ uncertainties, and above all, the combination of different probes, ranging from number counts to three-point statistics. This will improve the precision and accuracy of parameter estimation in the $\Lambda$CDM $\Omega_m-\sigma_8$ plane and in dynamical dark energy $w_0-w_\mathrm{a}$ extensions.
\begin{figure}[htbp]
\centering
\includegraphics[width=0.5\textwidth]{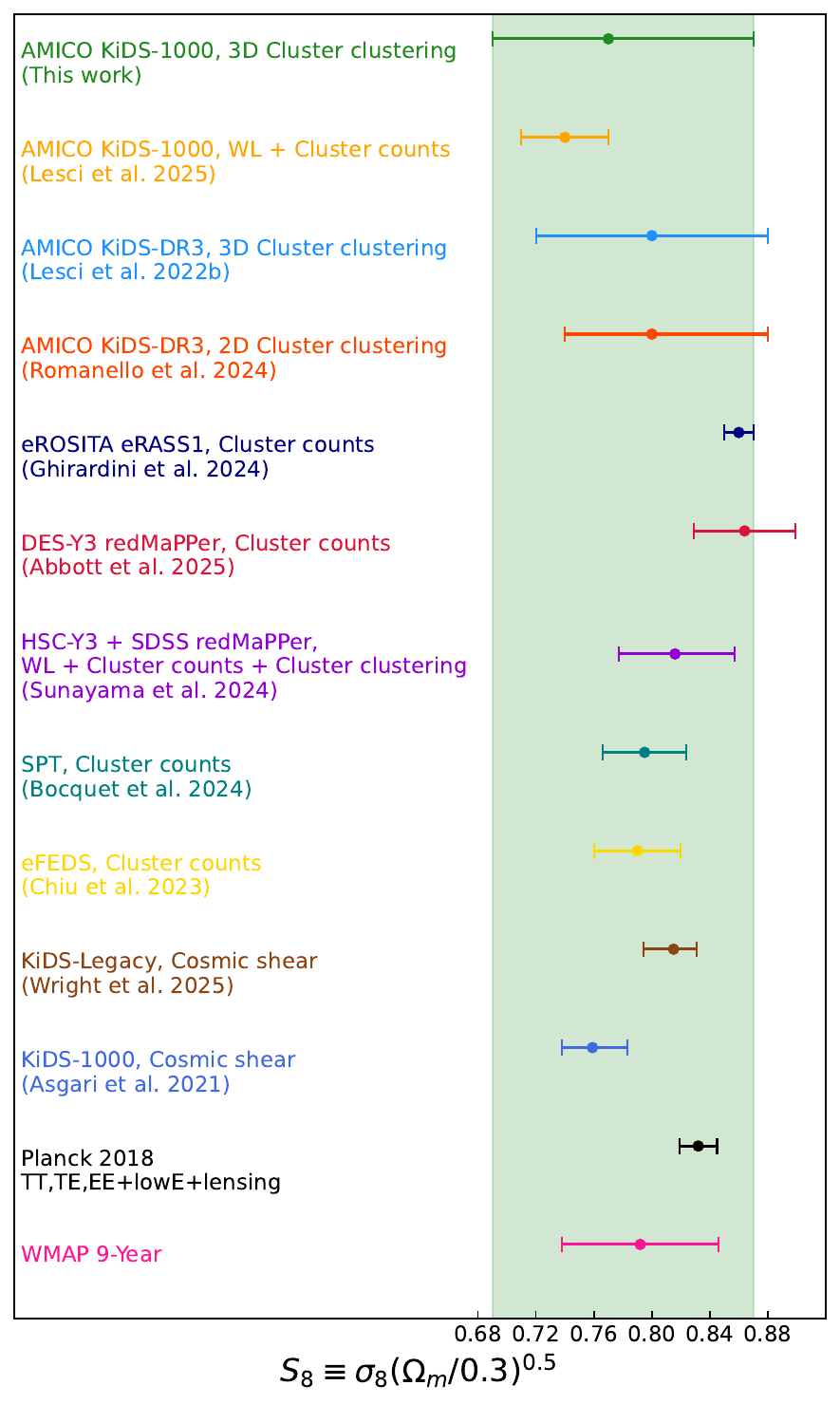}
\caption{Comparison of the constraints on the structure growth parameter, $S_8$. Dots represent the posterior median, the $16$-th and the $84$-th percentiles are delimited by the bars. From top to bottom: 3D
cluster clustering in the AMICO KiDS-1000 catalogue obtained in this work (green), with the green-shaded band that reproduces its uncertainty for a visual comparison; AMICO KiDS-1000 joint weak lensing and number counts from \citet[][gold]{Lesci25}; AMICO KiDS-DR3 3D cluster clustering from \citet[][pale blue]{Lesci22}; AMICO KiDS-DR3 angular cluster clustering from \citet[][dark orange]{Romanello24};
eROSITA All-Sky Survey number counts from \citet[][navy blue]{Ghirardini24}; 
DES cluster counts from \citet[][red]{Abbott25}; HSC plus SDSS joint modelling of weak lensing, number counts and clustering from \citet[][purple]{Sunayama24}; SPT cluster counts from \citet[][teal]{Bocquet24}; and eROSITA eFEDS from \citet[][yellow]{Chiu23}. Finally, we list measurements not based on cluster cosmology, such as KiDS-Legacy and KiDS-1000 Cosmic Shear constraints from \citet[][brown]{Wright25} and \citet[][royalblue]{Asgari21}; CMB results from (Planck+18, black) and WMAP \citep[][magenta]{WMAP13}.}
\label{Fig_comparison_S8}
\end{figure}

\section{Conclusions}
\label{Conclusions}
In this paper, we have presented the cosmological constraints on $\Omega_\mathrm{m}, \sigma_8$ and the derived parameter $S_8$, as obtained from the analysis of the two-point correlation function multipoles in the KiDS-1000 cluster data set. Our sample consisted of 9049 optically-selected galaxy clusters with $S/N \geq 3.5$, detected through the AMICO algorithm and distributed across two patches: one equatorial (KiDS-N) and one covering the South Galactic Pole (KiDS-S), for a total, effective area of 838.78 deg$^2$. \\
\indent We measured the monopole, quadrupole, and hexadecapole of the two-point correlation function over a variable range of scales from 5 $h^{-1}$ Mpc to 150 $h^{-1}$ Mpc, assuming the conservative richness cut from the weak-lensing analysis presented in \citet{Lesci25}. The catalogue spans a redshift range of $0.1\leq z<0.8$, and was divided into three different bins. Clusters have been selected according to their intrinsic richness: $\lambda^*_\mathrm{obs}\geq 20,25,30$ in $0.10\leq z<0.30$, $0.30\leq z<0.45$ and $0.45\leq z<0.80$, respectively. Errors were estimated through 500 bootstrap resamplings. \\
\indent We modelled the two-point correlation function, taking into account the main uncertainties affecting the clustering signal. For this purpose, we adopted the photo-$z$ uncertainty calibrated in \citet{Maturi25} using the KiDS cluster member galaxies that have a spectroscopic counterpart in the GAMA survey \citep{Driver11, Liske15}, given by $\sigma_z = 0.014(1+z_\mathrm{tr})$. This constitutes the main contribution to the small-scale damping of the monopole, and to the boost of the quadrupole and hexadecapole. Similarly, we assessed the impact of the redshift discretisation induced by the AMICO algorithm, as well as the one of velocity dispersion on the small-scale clustering signal, namely the FoG effect. We found that the former causes instability in the cluster clustering, especially at higher multipoles, because the data-data, random-random and data-random pairs only appear and are counted at discrete redshift separation steps, while the latter is subdominant and can be safely neglected. Furthermore, the clustering model was also corrected for geometric distortions. \\
\indent Deviations of $\lambda^*_\mathrm{obs}$ from $\lambda^*_\mathrm{tr}$ have been evaluated in \citet{Lesci25}, using AMICO KiDS mock catalogues produced by the \textsc{SinFoniA} code, along with the completeness and purity of the detections. Moreover, in order to relate the mass of the galaxy cluster to the intrinsic richness measured by AMICO, we adopted the mass-richness scaling relation calibrated in \citet{Lesci25}, using weak lensing and cluster number counts. These quantities are fundamental to estimate the effective bias of the cluster sample, for which we used the parametrisation provided by \citet{Tinker10}. In this case, switching from the halo mass function presented in \citet{Tinker08} to the fitting formulae of \citet{Despali16} leads to only modest variations, while leaving the overall result stable. \\
\indent From the MCMC Bayesian analysis, we found $\Omega_\mathrm{m}=0.24^{+0.02}_{-0.02}$ and $\sigma_8=0.86^{+0.13}_{-0.10}$, in perfect agreement with those obtained using the same AMICO KiDS-1000 galaxy cluster catalogue by \citet{Lesci25}, who also got evidence of a low dark matter density in the Universe. As shown in Fig. \ref{Fig_corner}, a $1\sigma$ level of agreement for the parameter $\sigma_8$ was also shared with AMICO KiDS-DR3 results from 3D and angular clustering. Notably, the FoM has increased by a factor of 1.8, with respect to the previous clustering analyses. This improvement was primarily driven by the reduced uncertainty on $\Omega_\mathrm{m}$. On the other hand, a $2\sigma$ tension is reported in $\Omega_\mathrm{m}$, with respect to the AMICO KiDS-DR3 $w(\theta)$ and Planck+18 CMB data. The robustness of our results is reinforced by the substantial overlap between the posterior distributions of all three redshift bins, individually considered, and is further confirmed when a more stringent $\lambda^*_\mathrm{obs}\geq30$ selection and a different binning strategy are applied. In addition, we verified that the effective bias of the cluster sample increases with the redshift bin, passing from $b_\mathrm{eff,1} = 2.40^{+0.38}_{-0.36}$, to $b_\mathrm{eff, 2}=2.89^{+0.49}_{-0.46}$, and finally to $b_\mathrm{eff,3}=3.68^{+0.64}_{-0.60}$, respectively, with relative uncertainties between $15\%$ and $17\%$.\\
\indent As a main outcome of this work, we found an independent constraint on the structure growth parameter, $S_8=0.77^{+0.10}_{-0.08}$. It shares a $1\sigma$ agreement with recent cosmological results reported in Fig. \ref{Fig_comparison_S8} and available in the literature. Most of them refer to galaxy cluster cosmology, in optical, millimetric and X-rays bands, but the accordance is kept also with respect to the KiDS-1000 and KiDS-Legacy cosmic shear findings \citep{Asgari21, Wright25}, as well as with CMB results from \citet{WMAP13} and Planck+18. \\
\indent Our work has shown that the cluster two-point correlation function can provide valuable constraints from Stage-III photometric-redshift surveys. Furthermore, it emphasises the relevance of probe combination, with cluster abundances and higher order statistics, in order to fully exploit the cosmological information contained in optically-selected catalogues. In the future, we expect improvements in both precision and accuracy of cosmological constraints, from the analysis of KiDS-Legacy \citep{Wright24}, which covers a final area of approximately $1350$ deg$^2$ of the KiDS/VIKING footprint, with a substantially deeper and larger redshift calibration sample. The inclusion of higher-order multipoles, as well as the models and methods developed here and their eventual extension from the $\Lambda$CDM to alternative cosmologies, will constitute a valuable testing ground for next-generation photometric redshift surveys, such as LSST \citep{MandelbaumLSST, Ivezic19LSST} and the \textit{Euclid} mission \citep{Laureijs11, Sartoris16, Scaramella22, Mellier25}.
%
\begin{acknowledgements}
     We acknowledge the financial contribution from the grant ASI n. 2024-10-HH.0 “Attività scientifiche per la missione \textit{Euclid} – fase E” and the use of computational resources from the parallel computing cluster of the Open Physics Hub (\url{https://site.unibo.it/openphysicshub/en}) at the Physics and Astronomy Department in Bologna. Based on observations made with ESO Telescopes at the La Silla Paranal Observatory under programme IDs 177.A-3016, 177.A-3017, 177.A-3018 and 179.A-2004, and on data products produced by the KiDS consortium. The KiDS production team acknowledges support from: Deutsche Forschungsgemeinschaft, ERC, NOVA and NWO-M grants; Target; the University of Padova, and the University Federico II (Naples). M.S. acknowledges financial contributions from contract INAF mainstream project 1.05.01.86.10, INAF Theory Grant 2023: Gravitational lensing detection of matter distribution at galaxy cluster boundaries and beyond (1.05.23.06.17), INAF Guest Observer Grant 2024: Towards anchoring the mass scale of galaxy clusters with galaxy kinematics (1.05.24.02.15), and contract Prin-MUR 2022 supported by Next Generation EU (n.20227RNLY3, The concordance cosmological model: stress-tests with galaxy clusters). This project has received funding from the European Union Horizon Europe Research and Innovation Action under grant agreement no. 101183153 -WST. Views and opinions expressed are however those of the author(s) only and do not necessarily reflect those of the European Union or the European Research Executive Agency (REA). Neither the European Union nor the REA can be held responsible for them. H. Hildebrandt is supported by a DFG Heisenberg grant (Hi 1495/5-1), the DFG Collaborative Research Center SFB1491, an ERC Consolidator Grant (No. 770935), and the DLR project 50QE2305. SJ acknowledges the Ramón y Cajal Fellowship (RYC2022--036431--I) from the Spanish Ministry of Science.
\end{acknowledgements}

%
%

\citestyle{aa}
\bibliographystyle{aa}
\bibliography{bibliografia}
%




   
  



\end{document}